\documentclass[12pt,english]{article}
\usepackage[T1]{fontenc}
\usepackage[latin1]{inputenc}
\usepackage{amssymb}

\makeatletter

\newlength{\lp}
\usepackage{amsfonts}

\usepackage{babel}
\makeatother
\begin{document}

\title{Via Aristotle, Leibniz \& Mach to a relativistic relational gravity }

\author{D. F. Roscoe}

\maketitle
\begin{abstract}
In previous work we have shown how a worldview that has its origins
in the ideas of Aristotle, Leibniz and Mach leads to a quasi-classical
(that is, \emph{one-clock}) metric theory of gravitation (astro-ph/0107397)
which, for example, when applied to model low surface brightness spirals
(astro-ph/0306228), produces results that have, hitherto, only been
matched by Milgrom's MOND algorithm. In this paper we show how the
natural generalization of this worldview into a properly relativistic
two-clock theory, applied to model a spherically symmetric gravitational
source, produces results that cannot be distinguished from the canonical
picture for all the standard local tests and which, when interpreted
as a radiation model, produces no dipole radiation. Furthermore, although
black-holes within this picture have an event horizon at the usual
Schwarzschild radius, they \emph{do not} have an essential singularity
at the origin - the solutions are perfectly regular there. 
\end{abstract}

\section{Introduction}

\subsection{A brief history of ideas of space and time}

The conception of space as the container of material objects is generally
considered to have originated with Democritus and, for him, it provided
the stage upon which material things play out their existence - \emph{emptiness}
exists and is that which is devoid of the attribute of \emph{extendedness}
(although, interestingly, this latter conception seems to contain
elements of the opposite view upon which we shall comment later).
For Newton \cite{key-90}, an extension of the Democritian conception
was basic to his mechanics and, for him:

\begin{quote}
\emph{... absolute space, by its own nature and irrespective of anything
external, always remains immovable and similar to itself. }
\end{quote}
Thus, the absolute space of Newton was, like that of Democritus, the
stage upon which material things play out their existence - it had
an \emph{objective existence} for Newton and was primary to the order
of things. In a similar way, time - \emph{universal time,} an absolute
time which is the same everywhere - was also considered to possess
an objective existence, independently of space and independently of
all the things contained within space. The fusion of these two conceptions
provided Newton with the reference system - \emph{spatial coordinates}
defined at a \emph{particular time} - by means of which, as Newton
saw it, all motions could be quantified in a way which was completely
independent of the objects concerned. It is in this latter sense that
the Newtonian conception seems to depart fundamentally from that of
Democritus - if \emph{emptiness} exists and is devoid of the attribute
of \emph{extendedness} then, in modern terms, the \emph{emptiness}
of Democritus can have no \emph{metric} associated with it. But it
is precisely Newton's belief in \emph{absolute space \& time} (with
the implied virtual clocks and rods) that makes the Newtonian conception
a direct antecedant of Minkowski spacetime - that is, of an empty
space and time within which it is possible to have an internally consistent
discussion of the notion of \emph{metric.}

The contrary view is generally considered to have originated with
Aristotle \cite{key-107,key-89} for whom there was no such thing
as a \emph{void} - there was only the \emph{plenum} within which the
concept of the \emph{empty place} was meaningless and, in this, Aristotle
and Leibniz \cite{key-85} were at one. It fell to Leibniz, however,
to take a crucial step beyond the Aristotolian conception: in the
debate of Clarke-Leibniz (1715$\sim$1716) \cite{key-84} in which
Clarke argued for Newton's conception, Leibniz made three arguments
of which the second was:

\begin{quote}
\emph{Motion and position are real and detectable only in relation
to other objects ... therefore empty space, a void, and so space itself
is an unnecessary hypothesis.}
\end{quote}
That is, Leibniz introduced a \emph{relational} concept into the Aristotolian
world view - what we call \emph{space} is a projection of \emph{relationships}
between material bodies into the perceived world whilst what we call
\emph{time} is the projection of ordered \emph{change} into the perceived
world. Of the three arguments, this latter was the only one to which
Clarke had a good objection - essentially that \emph{accelerated motion,}
unlike uniform motion, can be percieved \emph{without} reference to
external bodies and is therefore, he argued, necessarily percieved
with respect to the \emph{absolute space} of Newton. It is of interest
to note, however, that in rebutting this particular argument of Leibniz,
Clarke, in the last letter of the correspondence, put his finger directly
upon one of the crucial consequences of a relational theory which
Leibniz had apparently not realized (but which Mach much later would)
stating as absurd that:

\begin{quote}
... \emph{the parts of a circulating body (suppose the sun) would
lose the} vis centrifuga \emph{arising from their circular motion
if all the extrinsic matter around them were annihilated.}
\end{quote}
This letter was sent on October 29th 1716 and Leibniz died on November
14th 1716 so that we were never to know what Leibniz's response might
have been.

Notwithstanding Leibniz's arguments against the Newtonian conception,
nor Berkeley's contemporary criticisms \cite{key-86}, which were
very similar to those of Leibniz and are the direct antecedants of
Mach's, the practical success of the Newtonian prescription subdued
any serious interest in the matter for the next 150 years or so until
Mach himself picked up the torch. In effect, he answered Clarke's
response to Leibniz's second argument by suggesting that the \emph{inertia}
of bodies is somehow induced within them by the large-scale distribution
of material in the universe:

\begin{quote}
\emph{... I have remained to the present day the only one who insists
upon referring the law of inertia to the earth and, in the case of
motions of great spatial and temporal extent, to the fixed stars ...}
\cite{key-78}
\end{quote}
thereby generalizing Leibniz's conception of a relational universe.
Mach was equally clear in expressing his views about the nature of
time: in effect, he viewed \emph{time} (specifically Newton's \emph{absolute
time}) as a meaningless abstraction. All that we can ever do, he argued
in \cite{key-78}, is to measure \emph{change} within one system against
\emph{change} in a second system which has been defined as the standard
(eg it takes half of one complete rotation of the earth about its
own axis to walk thirty miles).

Whilst Mach was clear about the origins of inertia (in the fixed stars),
he did not hypothesize any mechanism by which this conviction might
be realized and it fell to others to make the attempt - a typical
(although incomplete) list might include the names of Einstein \cite{key-87},
Sciama \cite{key-83}, Hoyle \& Narlikar and Sachs \cite{key-81,key-82}
for approaches based on canonical ideas of spacetime, and the names
of Ghosh \cite{key-80} and Assis \cite{key-79} for approaches based
on quasi-Newtonian ideas.

It is perhaps one of the great ironies of 20thC science that Einstein,
having coined the name \emph{Mach's Principle} for Mach's original
suggestion and setting out to find a theory which satisfied the newly
named Principle, should end up with a theory which, whilst albiet
enormously successful, is more an heir to the ideas of Democritus
and Newton than to the ideas of Aristotle and Leibniz. One only has
to consider the special case solution of Minkowski spacetime, which
is empty but metrical, to appreciate this fact.

\section{From Leibniz to inertia as a relational property}

In our original paper on this topic \cite{key-3}, we took the general
position of Leibniz, about the relational nature of space, to be self-evident
and considered the question of \emph{metric} within this general conceptualization.
Briefly, we began by constructing a model universe populated by elementary
particles whose only property was that of \emph{countability.} We
were then able to arrive at a theory within which a metrical three-space
(generally non-Euclidean) is projected as a secondary construct out
of the relationships which hold within the primary universal distribution
of elementary material. The question of how \emph{time} arose within
the theory is particularly interesting: the simple requirement that
\emph{time} should be defined within the theory in such a way that
Newton's Third Law was automatically satisfied had the direct consequence
that \emph{time} became an explicit measure of change within the system,
very much as anticipated by Mach. The overall result was a quasi-classical
(one-clock) theory of relational gravitation within which:

\begin{itemize}
\item conditions of global dynamical equilibrium (that is, everywhere inertial)
held if the elementary material was distributed fractally, $D=2$; 
\item point-source perturbations of the $D=2$ distribution recovered the
usual Newtonian prescriptions for gravitational effects; 
\item cylindrically symmetric perturbations of the $D=2$ distribution,
used as rudimentary models of spiral galaxies and applied to model
the dynamics of low surface brightness galaxies \cite{key-12}, gave
results which were directly comparable in quality to those obtained
from Milgrom's MOND algorithm \cite{key-61,key-62,key-63} - and far
superior to anything achieved by the multi-parameter CDM models. 
\end{itemize}
The first of these three points refers to the universe that Leibniz
was effectively considering in his debate with Clarke of 1715$\sim$1716
- one within which inertial effects play no part. The second and third
refer to the universe that Clarke used to refute Liebniz's second
argument, and the one that Mach had in mind - the universe of rotations
and accelerations. Thus, given our original Leibnizian worldview,
we see that inertial effects themselves have their fundamental source
in changed material relationships - they, too, are \emph{relational}
in nature.

\subsection{A modern Le Sage theory}

An unforeseen consequence of the approach of \cite{key-3} was that
the distribution of elementary material from which the $D=2$ equilibrium
universe was constructed turned out to have properties more properly
associated with ideas of a \emph{material vacuum} rather than a distribution
of ordinary material as we conventionally understand it.

To be specific, each of the particles in the equilibrium universe
has an arbitrarily directed \emph{psuedo-velocity} property associated
with it, the magnitude of which is a global constant (the same for
every particle). However, this psuedo-velocity property is not a velocity
in the conventional sense, but is more accurately thought of as being
a conversion factor from length scales to time scales - very much
like Bondi's interpretation of $c$, the light speed. Thus, one can
can think of the matter distribution in the equilibrium universe as
similar to a \emph{photon gas} with the Bondian complexity that this
`gas' arbitrates between length scales and time scales - within the
context of the theory, this is a direct consequence of the fact that
\emph{time} arises automatically as a measure of \emph{change} within
the particle distribution very much as Mach conceived it.

Gravitational processes within this universe arise when the $D=2$
equilibrium distribution of the material vacuum is disturbed by discrete
distributions of ordinary material. Although the theory is cauched
in terms of a conventional metric description with equations of motion
arising in the canonical fashion from a variational principle, it
can also be thought of as a modern Le Sage-type theory \cite{key-75}
(in French, but see also Aronson \cite{key-76} for a discussion in
English ) - that is, as a \emph{pushing gravity} theory within which
gravitational effects arise as a consequence of momentum transfer
between the ever-moving particles of the vacuum gas and ordinary ponderable
material. A modern review of the history of such theories can be found
in \cite{key-94}.

The property of inertia itself (resistance to change in motion) can
be understood in the same way so that obvious similarities with the
ideas of Haisch, Rueda \cite{key-91} and Haisch, Rueda \& Puthoff
\cite{key-92} (concerning the origin of inertia in the electromagnetic
zero-point field) can be discerned.

\section{The relativistic generalization}

The basic purpose of this paper is to show how our Leibnizian worldview
of a relational universe can be generalized to the case of relativistic
gravitation. Surprisingly, we are able to show that, for the general
$N$-body case, linear momentum is exactly conserved and, in the particular
case of modelling point-source gravitation, the resulting model is
shown to:

\begin{itemize}
\item pass all of the standard local tests; 
\item have the usual one-way event horizon at the usual Schwarzschild radius,
$R=R_{S}$ say; 
\item have \emph{no singularity} at $R=0$ - instead, what happens is that
inside some boundary $R=R_{0}<R_{S}$ gravitation reverses its signature
so that, once a particle crosses the black-hole boundary $R=R_{S}$
then it orbits in an oscillatory fashion about the interior $R=R_{0}$
boundary; 
\item be explicity without dipole radiation. 
\end{itemize}

\subsection{The general argument}

In order to understand, in particular, how linear momentum is conserved
for the $N$-body case, it is useful to repeat, in a much-compacted
and simplified form, the general argument of \cite{key-3} modified
to account for the relativistic generalization.

Following in the tradition of Aristotle, Leibniz, Berkeley and Mach
we argue that no consistent cosmology should admit the possibility
of an internally consistent discussion of \emph{empty} metrical space
\& time - unlike, for example, General Relativity which has the empty
spacetime of Minkowski as a particular solution. Recognizing that
the most simple space \& time to visualize is one which is everywhere
inertial, then our worldview was distilled into the question:

\begin{quote}
\emph{Is it possible to conceive a globally inertial space \& time
which is irreducibly associated with a non-trivial global mass distribution
and, if so, what are the properties of this distribution?}
\end{quote}
The primary step taken in answer to the question was the recognition
that, on large enough scales in the universe of our experience (say
$>30\, Mpc$), the amount of matter in a given spherical volume in
a given epoch can be considered as a well-defined (monotonic) function
of the sphere's radius.

\begin{quote}
\emph{It follows immediately that, within the universe of our experience,
we can choose to} \textbf{\emph{define}} \emph{the radius of any given
sphere at any given epoch in terms of the amount of material contained
within it. }
\end{quote}
We now use this idea to construct a rudimentary model universe which
is isotropic at all points (so that rotational invariance can be assumed
everywhere), is populated entirely by identical particles having only
the property of \emph{countability} and within which the radius of
any spherical volume at any given time is defined in terms of the
amount of mass contained within the volume: \begin{equation}
R=f(Nm_{0},t)\rightarrow\delta R\approx f(Nm_{0}+\delta N\, m_{0},t)-f(Nm_{0},t)\label{eqn0}\end{equation}
 where $N$ is the number of particles concerned, $m_{0}$ is a scaling
constant having dimensions of mass and $f$ is an arbitrary monotonic
increasing function of $N$. In this way, we have immediately defined
one particular invariant linear measurement (a radial one) such that
it becomes undefined in the absence of matter - in effect, we have
provided an uncalibrated metric (since $f$ is undefined) which follows
Leibniz in the required sense for any displacement which is purely
radial\emph{.}

We now look for ways of generalizing this idea so that we can assign
a metric to \emph{arbitrary} displacements within the model universe.
To this end, we consider the first of (\ref{eqn0}) to be primary
and then invert it to give the uncalibrated mass model, \[
Mass\equiv Nm_{0}=\mathcal{{\mathcal{M}}}(R,t)\equiv{\mathcal{M}}(x^{1},x^{2},x^{3},x^{4}),\]
 of our rudimentary model universe (here, $x^{4}\equiv ct$). Note
that we make no assumptions about the relation of the spatial coordinates,
$(x^{1},x^{2},x^{3})$, to the radial displacement, $R$. Now consider
the normal gradient vector $n_{a}=\nabla_{a}{\mathcal{M}}$ (for which
choice a detailed physical argument is given in \cite{key-3}) and
the change in this arising from a displacement $dx^{k}$, \begin{equation}
dn_{a}=\nabla_{i}\left(\nabla_{a}{\mathcal{\mathcal{M}}}\right)\, dx^{i}\,,\label{eqn0A}\end{equation}
 where we assume that the geometrical connections required to give
this latter expression an unambiguous meaning are the usual metrical
connections - except of course, the metric tensor $g_{ab}$ is not
yet defined.

Given that $\nabla_{a}\nabla_{b}{\mathcal{\mathcal{M}}}$ is nonsingular,
then (\ref{eqn0A}) provides a 1:1 mapping between the contravariant
vector $dx^{a}$ and the covariant vector $dn_{a}$ so that, in the
absence of any other definition, we can \emph{define} $dn_{a}$ to
be the covariant form of $dx^{a}$. In this latter case the metric
tensor automatically becomes $g_{ab}\equiv\nabla_{a}\nabla_{b}{\mathcal{\mathcal{M}}}$
which, through the implied metrical connections, is a highly non-linear
equation defining $g_{ab}$ to within the specification of ${\mathcal{\mathcal{M}}}$.
Note that, at this stage, ${\mathcal{\mathcal{M}}}$ is assumed to
be arbitrary (beyond the basic requirement of monotonicity) which
is equivalent to saying that the linear scale is \emph{uncalibrated.}
Thus, assuming the usual weak equivalence principle which requires
that the action\[
{\mathcal{\mathcal{I}}}=\int_{P_{0}}^{P_{1}}\sqrt{g_{ij}\,\dot{x}^{i}\dot{x}^{j}}\, ds\]
 is minimized for gravitational trajectories, then:

\begin{quote}
\emph{the equations of motion will be invariant} \emph{under arbitrary
(monotonic) transformations of the linear scale up to, but not including,
the specification of ${\mathcal{\mathcal{M}}}$.}
\end{quote}
We now ask what form ${\mathcal{\mathcal{M}}}$ must have so that
flat spacetime, $g_{ab}=\gamma_{ab}$, is recovered. The question
is trivially answered giving \[
\mathcal{{\mathcal{M}}}(R,ct)=\left(x^{1}\right)^{2}+\left(x^{2}\right)^{2}+\left(x^{3}\right)^{2}+\left(x^{4}\right)^{2}\equiv R^{2}-c^{2}t^{2}.\]
 The corresponding result from the quasi-classical analysis of \cite{key-3}
(for the existence of Euclidean frames) was that $\mathcal{{\mathcal{M}}}(R)=R^{2}$,
valid about any centre so that mass is distributed fractally with
$D=2$ in that case. In the present case, for which ${\mathcal{\mathcal{M}}}(R,ct)=R^{2}-c^{2}t^{2}$,
we can say that the mass distribution tends to become fractal, $D=2$,
when very large spatial scales are considered over very short time
scales - that is, for $ct<<R$.

Finally, we show, in appendix \ref{app:The-Material-Vacuum}, how
this definition of $\mathcal{{\mathcal{M}}}(R,ct)$ can be considered
as a classical model of a rudimentary material vacuum which, because
it is associated with $g_{ab}=\gamma_{ab}$, we call the \emph{Minkowski
vacuum.} Thus, on large enough distance scales and short enough temporal
scales, the relativistic material vacuum also tends to fractal with
$D=2$.

\subsection{Fractal $D=2$ universe }

The original quasi-classical theory states that the material vacuum
of the equilibrium model is exactly fractal with $D=2$, whilst the
current relativistic theory states that it tends to this state on
large enough spatial scales and short enough temporal scales. If one
imagines that ordinary matter somehow `condenses' out of the material
vacuum (say by collision processes) then it follows that the distribution
of ordinary material must be fractal-like with \emph{maximum} fractal
dimension $D=2$ and where we could expect this latter figure to be
closely approached in a `maximally evolved' system.

The idea that the universe might, indeed, be fractal is an old one
which can be traced to Charlier's conception of an hierarchical universe
\cite{key-5,key-6,key-7}. The contemporary debate was probably initiated
by Pietronero in 1987 \cite{key-98} and Coleman et al in 1988 \cite{key-100}
and, in recent years, several quantitative analyses of both pencil-beam
and wide-angle surveys of galaxy distributions have been performed:
three recent examples are give by Joyce et al \cite{key-14} who analysed
the CfA2-South catalogue to find fractal behaviour with $D\,$=$\,1.9\pm0.1$;
Sylos Labini \& Montuori \cite{key-16} analysed the APM-Stromlo survey
to find fractal behaviour with $D\,$=$\,2.1\pm0.1$, whilst Sylos
Labini et al \cite{key-17} analysed the Perseus-Pisces survey to
find fractal behaviour with $D\,$=$\,2.0\pm0.1$. There are many
other papers of this nature in the literature all supporting the view
that, out to about $50Mpc$ at least, galaxy distributions appear
to be fractal with $D\approx2$. This latter view is now widely accepted
(for example, see Wu, Lahav \& Rees \cite{key-26}), and the open
question has become whether or not there is a transition to homogeneity
on some sufficiently large scale.

The argument has recently reduced to a question of statistics: basically,
the proponents of the fractal view argue that the statistical tools
(eg correlation function methods) widely used to analyse galaxy distributions
by the proponents of the opposite view are deeply rooted in classical
ideas of statistics and implicitly assume that the distributions from
which samples are drawn are homogeneous in the first place. Thus,
much effort is being expended developing tools appropriate to analysing
samples drawn from more general classes of populations - a general
focus being the idea that one should not discuss fractal structures
in terms of the correlation amplitude since the only meaningful quantity
is the exponent characterizing the fractal behaviour. Recent papers
arguing this general point of view are Sylos Labini \& Gabrielli \cite{key-105}
and Gabrielli \& Sylos Labini \cite{key-106}.

According to the present theory, the debate should be settled in favour
of the globally fractal universe, $D=2$.

\section{The emergence of gravitation}

It was shown in \cite{key-12} how Newtonian gravitation emerges as
a spherical perturbation of the equilibrium background, and a similar
result is true here, as we shall show in detail from \S\ref{sec:The-emergence-of}
onwards. Thus, for example, we suppose that a spherically symmetric
distribution of ponderable material can be represented as a spherically
symmetric disturbance of the vacuum so that, if ${\mathcal{\mathcal{M}}}^{(0)}$
represents the undisturbed vacuum, then\[
\mathcal{{\mathcal{M}}}=\mathcal{{\mathcal{M}}}^{(0)}+\epsilon\]
 represents the disturbed vacuum and\begin{eqnarray}
g_{ab} & = & \nabla_{a}\nabla_{b}\mathcal{{\mathcal{M}}}\equiv\frac{\partial^{2}\mathcal{{\mathcal{M}}}}{\partial x^{a}\partial x^{b}}-\Gamma_{ab}^{k}\frac{\partial\mathcal{{\mathcal{M}}}}{\partial x^{k}}\label{eqn3}\\
\nonumber \\\Gamma_{ab}^{k} & \equiv & \frac{1}{2}g^{kr}\left(\frac{\partial g_{br}}{\partial x^{a}}+\frac{\partial g_{ra}}{\partial x^{b}}-\frac{\partial g_{ab}}{\partial x^{r}}\right)\nonumber \end{eqnarray}
 gives the metric tensor in this disturbed vacuum.

\subsection{Linear momentum conservation within the formalism}

Suppose that we have a finite ensemble of ponderable mass particles,
all having \emph{non-relativtistic} velocities\emph{,} embedded in
the $D=2$ equilibrium background. Then, we can suppose that all discussion
of momentum conservation can be referred to the mass centre of the
ensemble, and that this mass centre is in dynamic equilibrium with
the background.

For any system of particles of masses $M_{1},...,M_{N}$, described
from a centre-of-mass frame, the integrated Newtonian momentum-conservation
equation becomes \[
M_{1}\textbf{R}^{1}+M_{2}\textbf{R}^{2}+...+M_{N}\textbf{R}^{N}=0.\]
 The masses appearing in this equation are now arbitrarily partitioned
into the pair of ensembles $M_{1},...,M_{k-1}$ and $M_{k},...,M_{N}$.
Defining the mass of the whole system as $M$, and the mass of the
ensemble $M_{1},...,M_{k-1}$ as $m$, then the foregoing equation
can be written as \[
m\textbf{r}+(M-m)\textbf{R}=0,\]
 where $\textbf{r}$ and $\textbf{R}$ are the respective mass-centres
of the two, arbitrarily defined, particle ensembles defined with respect
to the mass centre of the whole ensemble. Any interaction can then
be considered as being between the particle ensemble of mass $m$
and the rest of the ensemble, having mass $M-m$. Whatever the details
of this interaction, these two particle ensembles must, together,
evolve from their initial state in such a way that linear momentum
is conserved for all $t>0$ so that, always, \begin{equation}
m\textbf{r}=-(M-m)\textbf{R}\rightarrow\mathbf{r}=\lambda\mathbf{R}\label{GMC}\end{equation}
 But, it is easily shown that, for \emph{unspecified} $\mathcal{M}$,
the equations of motion arising from (\ref{eqn3}) are scale-invariant
under the light-cone preserving transformation $\textbf{r}=\lambda\textbf{R},\, t=\lambda T$
for non-zero constant $\lambda$; in particular, this is true for
the $\lambda$ value defined at (\ref{GMC}). It follows that the
equations of motion for $\textbf{r}$ and $t$ will transform into
\textit{identical} equations of motion for $\lambda\textbf{R}$ and
$\lambda T$ so that, with the initial condition $\textbf{r}(0)=\lambda\textbf{R}(0)$,
the calculated trajectories will satisfy $\textbf{r}=\lambda\textbf{R}$
for all time. That is, linear momentum is exactly conserved within
the formalism.

\section{Point-source disturbances in the Minkowski vacuum\label{sec:The-emergence-of}}

\label{sec.3} In the following, a point-mass gravitating source is
represented as a spherically symmetric point-source disturbance of
the Minkowski vacuum and, in appendix \ref{appA}, we show that a
first-order approximation to such a disturbance has the structure\begin{equation}
{\mathcal{M}}^{(1)}(\mathbf{x})={\mathcal{M}}^{(0)}(\mathbf{x})+\frac{H(R-ct)}{R},\label{eqn4}\end{equation}
 where $H$ is a twice-differentiable function. Following the details
of appendix \ref{appA}, we find that the corresponding perturbed
metric is given by\[
g_{ab}^{(1)}\equiv\gamma_{ab}+\epsilon_{ab}^{(1)}=\gamma_{ab}+\frac{\partial^{2}E^{(1)}}{\partial x^{a}\partial x^{b}}=\gamma_{ab}+\frac{\partial^{2}}{\partial x^{a}\partial x^{b}}\left(\frac{H}{R}\right).\]
 Expanding this expression, forming the corresponding proper-time
element, and transforming to spherical polar coordinates we obtain,
as detailed in appendix \ref{appB}, \begin{eqnarray}
d\tau^{2} & = & \Delta_{0}(R,dR,d\phi,dt)+\Delta_{1}(R,dR,d\phi,dt)+\Delta_{2}(R,dR,d\phi,dt)\label{eqn5}\\
\Delta_{0} & \equiv & c^{2}dt^{2}-dR^{2}-R^{2}d\Phi^{2},\nonumber \\
\Delta_{1} & \equiv & \frac{\ddot{H}}{R}\left(dR^{2}-2cdRdt+c^{2}dt^{2}\right),\nonumber \\
\Delta_{2} & \equiv & \frac{\dot{H}}{R^{2}}\left(-2dR^{2}+R^{2}d\phi^{2}+2cdRdt\right)-\frac{H}{R^{3}}\left(-2dR^{2}+R^{2}d\phi^{2}\right).\nonumber \end{eqnarray}
 The $\Delta_{0}$ component represents the infinitessimal proper-time
registered by a test-particle moving in the undisturbed Minkowski
vacuum, whilst $\Delta_{1}$ and $\Delta_{2}$ describe the perturbation
of this proper-time element caused by the point-source.

\subsection{The integrated effect of many disturbances}

Now suppose that the disturbance which gives rise to (\ref{eqn5})
is simply one of a continual train of identical disturbances passing
through the particle with a regular frequency measured in the rest-frame
of the source (the source is considered as an oscillator), then we
can replace $\ddot{H}$, $\dot{H}$ and $H$ by their mean-values
between successive minima in the train of disturbances, and redefine
$d\tau^{2}$ of (\ref{eqn5}) as a measure of the mean proper-time
registered by the test-particle between these successive minima. With
this understanding, and writing the mean-value of $\ddot{H}$ as $\lambda$,
then, to $O(1/R)$, a spherically symmetric disturbance in the Minkowski
vacuum passing through the test-particle causes the particle to register
a mean infinitessimal proper-time, given by \begin{equation}
d\tau^{2}\approx c^{2}\left(1+\frac{\lambda}{R}\right)dt^{2}-\frac{2c\lambda}{R}dRdt-\left(1-\frac{\lambda}{R}\right)dR^{2}-R^{2}d\phi^{2},\label{eqn5A}\end{equation}
 measured over the interval for which the disturbance can be said
to be passing through the particle. Now, since the integrated magnitude
of the deviation of the disturbed proper-time from the undisturbed
proper-time, measured over any finite interval, will depend on the
frequency with which the disturbances arrive at the test-particle
then we can deduce that the undetermined parameter $\lambda$ must
be a measure of the frequency, and hence rest-mass, of the disturbing
source. Consequently, $\lambda=\beta M_{0}$, for constant $\beta$
and rest-mass $M_{0}$, and the proper-time element (\ref{eqn5A})
is now reduced to the general form of that arising in any conventional
metric-gravitation theory.

Finally, we notice that the general form of (\ref{eqn5A}) is identical
to that of the Eddington form of the Schwarzschild proper-time element
and that the two forms match exactly if $\lambda=-2\gamma M_{0}/c^{2}$;
consequently, under the transformation, \[
cdt\rightarrow cdt+\left(1-\frac{c^{2}R}{2\gamma M_{0}}\right)^{-1}dR\]
 then \[
d\tau^{2}\approx\Delta_{0}+\Delta_{1}\equiv c^{2}\left(1-\frac{2\gamma M_{0}}{c^{2}R}\right)dt^{2}-R^{2}d\phi^{2}-\left(1-\frac{2\gamma M_{0}}{c^{2}R}\right)^{-1}dR^{2},\]
 and this form is identical to the Schwarzschild proper-time element,
as required.

\section{Black holes without singularities}

The proper-time element, (\ref{eqn5}), is only fully determined when
the neglected $\dot{H}/R^{2}$ and $H/R^{3}$ terms are included.
It transpires that their presence \emph{removes} the essential singularity
at $R=0$ in the line element.

In the previous section, we defined $\lambda$ as the mean value of
$\ddot{H}$ between successive minima in the train of vacuum disturbances
generated by the central massive source. If we now define $\alpha$
and $\beta$, in a similar way, as the corresponding mean values of
$\dot{H}$ and $H$, and make the definitions \begin{eqnarray*}
X & = & 1+\frac{\lambda}{R}\\
Y & = & 1-\frac{\alpha}{R^{2}}-\frac{\beta}{R^{3}}\\
Z & = & \frac{\lambda}{R}-\frac{\alpha}{R^{2}},\end{eqnarray*}
 then the full line element, given at (\ref{eqn5}), leads fairly
easily to the equations of motion \begin{eqnarray*}
\frac{k_{1}}{R^{2}}\frac{dR}{d\Phi} & = & \pm\sqrt{\frac{Y(-k_{0}^{2}Y+XY+k_{1}^{2}X/R^{2})}{X^{2}-2XY-Z^{2}}}\equiv\pm W\\
\frac{d\Phi}{cdt} & = & \frac{Xk_{1}}{R^{2}\left(Yk_{0}\pm ZW\right)},\end{eqnarray*}
 so that the radial velocity equation is given by \begin{equation}
\frac{dR}{cdt}=\frac{dR}{d\Phi}\frac{d\Phi}{cdt}=\frac{\pm XW}{Yk_{0}\pm ZW}\label{eqn6}\end{equation}
 where the choice of signature `$\pm$' determines if radial motion
is towards, or away from, the gravitating source. Note that $\alpha=\beta=0$
yields the corresponding GR equations.

In the previous section, we showed $\lambda=-2\gamma M/c^{2}$ so
that the Schwarzschild radius is at $R_{s}=-\lambda$. Consequently,
from the definition of $X$, we see that $X=0$ in (\ref{eqn6}) on
the Schwarzschild boundary, and an analysis of the equation in the
region of this boundary shows that test-particles can cross it when
inward-bound (signature choice is `$-$'), but cannot cross it when
outward bound (signature choice is `$+$'). In this sense, of course,
the presented theory is in direct accord with the GR model.

The real differences between the presented theory and GR emerge when
the radial equation (\ref{eqn6}) is analysed in the limit of $R\rightarrow0$.
Specifically, we find \[
\frac{dR}{cdt}\approx\pm\left(\frac{\lambda}{\alpha}\right)R,\]
 so that, as an immediate consequence, \[
\frac{d^{2}R}{c^{2}dt^{2}}=\left(\frac{\lambda}{\alpha}\right)^{2}R\]
 which implies that gravitational attraction `turns off' at some $R=R^{*}$
for $0<R^{*}<R_{s}$, and becomes gravitational repulsion in the region
$0<R<R^{*}$. The origin, $R=0$, is a point of unstable equilibrium,
and so can be considered as the top of a `potential hill'. Consequently,
the essential singularity which exists in the GR model at $R=0$ does
not exist in the perturbed Minkowski vacuum model.

\section{The absence of dipole effects}

We have already noted that linear momentum is exactly conserved within
the formalism (at least for non-relativistic systems) so that there
can be no dipole effects arising from the non-conservation of linear
momentum. In the following, we show that dipole radiation in general
is \emph{explicitly} absent from the theory in its radiation model
interpretation.

We begin by noting that the far-field solution of any gravitationally
radiating binary system will necessarily be spherically symmetric.
Consequently, if we wish to determine whether or not dipole radiation
exists at all according to the theory, then it is sufficient to investigate
the general radiation solution for the spherically symmetric case
up to its lowest order multipole. The details of this analysis are
given in appendices \ref{appA} and \ref{appC} where it is found
that \[
{\mathcal{M}}(\underline{x})={\mathcal{M}}^{(0)}(\underline{x})+\frac{H(R-ct)}{R}+\frac{G(R-ct)}{R^{N}}\times(regular~function)+...,\]
 where $N>2$ so that dipole terms are explicitly excluded from the
general multipole solution. The corresponding metric is then found
by using the relationship\begin{eqnarray*}
g_{ab} & = & \left(\frac{\partial^{2}{\mathcal{M}}}{\partial x^{a}\partial x^{b}}-\Gamma_{ab}^{k}\frac{\partial{\mathcal{M}}}{\partial x^{k}}\right)\\
\\\Gamma_{ab}^{k} & \equiv & \frac{1}{2}g^{kr}\left(\frac{\partial g_{br}}{\partial x^{a}}+\frac{\partial g_{ra}}{\partial x^{b}}-\frac{\partial g_{ab}}{\partial x^{r}}\right).\end{eqnarray*}

\section{Conclusions}

\label{sec.8} Using a view of the universe that can be traced to
Aristotle, via Mach, Liebniz and Berkeley we have arrived at a relativistic
theory of \emph{relational gravity} which, apart from satisfying all
the standard local tests, conserves linear momentum exactly in a slow
moving $N$-body system and is explicitly free of any dipole gravitational
radiation component. We can therefore expect that it will be `good'
for the binary pulsar tests also (eg Will \cite{4}). Furthermore,
the theory predicts the existence of blackholes with the usual Schwarzschild
radius but \emph{without} the essential singularities that exist in
canonical theory.

\appendix

\section{The Material Vacuum\label{app:The-Material-Vacuum}}

It is easily shown that\begin{eqnarray*}
{\mathcal{M}}^{(0)} & \equiv & \frac{1}{2}(x^{i}-x_{0}^{i})(x^{j}-x_{0}^{j})\gamma_{ij}\equiv\Delta_{0}\end{eqnarray*}
 for an arbitrary origin $x_{0}^{i}$ which satisfies $\Delta_{0}>0$
gives rise directly, via (\ref{eqn3}), to $g_{ab}=\gamma_{ab}$.
A natural generalization of ${\mathcal{M}}^{(0)}$, obtained by summing
over all possible origins, is given by\begin{equation}
{\mathcal{M}}^{(0)}=\sum_{x_{0}^{1}}\sum_{x_{0}^{2}}\sum_{x_{0}^{3}}\sum_{x_{0}^{4}}\alpha(x_{0}^{1},x_{0}^{2},x_{0}^{3},x_{0}^{4})\Delta_{0}\label{eqn.Matvac3}\end{equation}
 where $\alpha(x_{0}^{1},x_{0}^{2},x_{0}^{3},x_{0}^{4})$ is such
that ${\mathcal{M}}^{(0)}$ also gives $g_{ab}=\gamma_{ab}$ when
put into (\ref{eqn3}).

An understanding of the meaning of this ${\mathcal{M}}^{(0)}$ can
be had by considering the expanding surface \[
(x^{i}-x_{0}^{i})(x^{j}-x_{0}^{j})\gamma_{ij}\equiv(\textbf{x}-\textbf{x}_{0})^{2}-c^{2}(t-t_{0})^{2}=k^{2},\]
 for $k$ some real constant, generated by a single term of (\ref{eqn.Matvac3}).
It is easily shown that the radial speed of such an expanding surface
increases from $0$ to $c$ on the range $|k|\leq|\textbf{x}-\textbf{x}_{0}|<\infty$.
It follows that ${\mathcal{M}}^{(0)}$, which is defined by (\ref{eqn.Matvac3})
at the spacetime point $(\textbf{x},ct)$ by summing over all admissible
origins $(\textbf{x}_{0},ct_{0})$, is a sum over an infinity of instantaneously
intersecting surfaces expanding from all possible directions and at
all possible subluminal speeds. In this way, we generate a classical
image of a continually fluctuating relativistically invariant material
vacuum. Consequently, (\ref{eqn.Matvac3}) can be interpreted as a
rudimentary model of a classical fluctuating material vacuum.

\section{Details of the Monopole Approximation}

\label{appB} In this appendix we give the details of the algebra
leading up to (\ref{eqn5}). Defining the parameter $S\equiv R-ct$
where $R^{2}=(x^{1})^{2}+(x^{2})^{2}+(x^{3})^{2}$, and using the
notation $x^{0}=ct$, then we find \begin{eqnarray}
\epsilon_{ab}^{(1)} & \equiv & \frac{\partial^{2}}{\partial x^{a}\partial x^{b}}\left(\frac{H}{R}\right)\nonumber \\
 & = & \frac{\ddot{H}}{R}S_{a}S_{b}+\frac{\dot{H}}{R^{2}}\left(RS_{ab}-R_{a}S_{b}-R_{b}S_{a}\right)-\frac{H}{R^{3}}\left(RR_{ab}-2R_{a}R_{b}\right),\label{eqn19}\end{eqnarray}
 where \[
\dot{H}\equiv\frac{dH}{dS};\]
 \[
S_{a}\equiv\frac{\partial S}{\partial x^{a}}=\frac{x_{a}(1-\delta_{a0})}{R}-\delta_{a0};\]
 \[
R_{a}\equiv\frac{\partial R}{\partial x^{a}}=\frac{x_{a}(1-\delta_{a0})}{R};\]
 \[
~S_{ab}\equiv\frac{\partial^{2}S}{\partial x^{a}\partial x^{b}}=\frac{(1-\delta_{a0})(1-\delta_{b0})}{R}\left(\delta_{ab}-\frac{x^{a}x^{b}}{R^{2}}\right)\]
 \[
~R_{ab}\equiv\frac{\partial^{2}R}{\partial x^{a}\partial x^{b}}\equiv S_{ab}.\]

Now consider a particle, having coordinates $(x^{a})$ relative to
the spatial origin, being displaced to $(x^{a}+dx^{a})$. The corresponding
proper-time element for this displacement is given by \[
d\tau^{2}=(\gamma_{ij}+\epsilon_{ij}^{(1)})dx^{i}dx^{j},\]
 so that, using (\ref{eqn19}), we find \begin{eqnarray*}
d\tau^{2} & = & \gamma_{ij}dx^{i}dx^{j}+\frac{\ddot{H}}{R}\left[\frac{1}{R}\left(x^{1}dx^{1}+x^{2}dx^{2}+x^{3}dx^{3}\right)-dx^{0}\right]^{2}\\
 & + & \frac{\dot{H}}{R^{2}}\left[(dx^{1})^{2}+(dx^{2})^{2}+(dx^{3})^{2}-\frac{3}{R^{2}}\left(x^{1}dx^{1}+x^{2}dx^{2}+x^{3}dx^{3}\right)^{2}\right.\\
 & + & \left.\frac{2}{R}\left(x^{1}dx^{1}+x^{2}dx^{2}+x^{3}dx^{3}\right)dx^{0}\right]\\
 & - & \frac{H}{R^{3}}\left[(dx^{1})^{2}+(dx^{2})^{2}+(dx^{3})^{2}-\frac{3}{R^{2}}\left(x^{1}dx^{1}+x^{2}dx^{2}+x^{3}dx^{3}\right)^{2}\right].\end{eqnarray*}
 Using the spherical polar coordinate transformations $x^{1}=R\cos\phi$,
$x^{2}=R\sin\phi$, $x^{3}=0\equiv\theta=\pi/2$ and $x^{0}=ct$,
then this expression can be written as \begin{eqnarray*}
d\tau^{2} & = & c^{2}dt^{2}-dR^{2}-R^{2}d\phi^{2}+\frac{\ddot{H}}{R}\left(dR^{2}-2cdRdt+c^{2}dt^{2}\right)\\
 & + & \frac{\dot{H}}{R^{2}}\left(-2dR^{2}+R^{2}d\phi^{2}+2cdRdt\right)-\frac{H}{R^{3}}\left(-2dR^{2}+R^{2}d\phi^{2}\right),\end{eqnarray*}
 for the result.

\section{Multi-pole disturbances in the vacuum}

\label{appA} The general formalism is given by \[
g_{ab}=\left(\frac{\partial^{2}{\mathcal{M}}}{\partial x^{a}\partial x^{b}}-\Gamma_{ab}^{k}\frac{\partial{\mathcal{M}}}{\partial x^{k}}\right)\]
 where \[
\Gamma_{ab}^{k}=\frac{1}{2}g^{kr}\left(\frac{\partial g_{br}}{\partial x^{a}}+\frac{\partial g_{ra}}{\partial x^{b}}-\frac{\partial g_{ab}}{\partial x^{r}}\right),\]
 and the Minkowski vacuum is given by \begin{eqnarray*}
{\mathcal{M}}(\mathbf{x}) & \equiv & {\mathcal{M}}^{(0)}(\mathbf{x})=\frac{1}{2}(x^{i}-x_{0}^{i})(x^{j}-x_{0}^{j})\gamma_{ij}\\
g_{ab} & = & \gamma_{ab}\end{eqnarray*}
 We can write a perturbed solution as \begin{eqnarray}
{\mathcal{M}}(\mathbf{x}) & = & {\mathcal{M}}^{(0)}(\mathbf{x})+E(\mathbf{x})\label{eqn7}\\
g_{ab} & = & \gamma_{ab}+\epsilon_{ab},\label{eqn8}\end{eqnarray}
 where $E(\mathbf{x})$ is a disturbance of the Minkowski vacuum and
$\epsilon_{ab}$ is the corresponding perturbation of the metric-tensor.
By writing \begin{eqnarray}
\Delta_{ab} & \equiv & \frac{\partial^{2}.}{\partial x^{a}\partial x^{b}}-\Gamma_{ab}^{k}\frac{\partial.}{\partial x^{k}}\label{eqn9}\\
\Box & \equiv & g^{ij}\Delta_{ij}\label{eqn10}\end{eqnarray}
 then the formalism can be expressed as \begin{equation}
g_{ab}=\Delta_{ab}{\mathcal{M}}(\mathbf{x})\label{eqn11}\end{equation}
 where\begin{equation}
\Box{\mathcal{M}}=4.\label{eqn12}\end{equation}
 Suppose we express (\ref{eqn7}) and (\ref{eqn8}) in the form \begin{eqnarray*}
{\mathcal{M}}(\mathbf{x}) & = & {\mathcal{M}}^{(0)}(\mathbf{x})+\sum_{r=1}^{\infty}E^{(r)}(\mathbf{x}),\\
g_{ab} & = & \gamma_{ab}+\sum_{r=1}^{\infty}\epsilon_{ab}^{(r)},\end{eqnarray*}
 where the summations converge uniformly to $E(\mathbf{x})$ and $\epsilon_{ab}$
respectively, then \begin{eqnarray}
{\mathcal{M}}^{(n)}(\mathbf{x}) & = & {\mathcal{M}}^{(0)}(\mathbf{x})+\sum_{r=1}^{n}E^{(r)}(\mathbf{x})\label{eqn13}\\
g_{ab}^{(n)} & = & \gamma_{ab}+\sum_{r=1}^{n}\epsilon_{ab}^{(r)}\label{eqn14}\end{eqnarray}
 can be considered as an n-th order approximation to the exact solution
and where, in this notation, ${\mathcal{M}}^{(0)}$ represents the
Minkowski vacuum and $g_{ab}^{(0)}\equiv\gamma_{ab}$. The (n+1)-th
order approximation \begin{eqnarray}
{\mathcal{M}}^{(n+1)} & = & {\mathcal{M}}^{(n)}+E^{(n+1)}\label{eqn15}\\
g_{ab}^{(n+1)} & = & g_{ab}^{(n)}+\epsilon_{ab}^{(n+1)};\,\, n=0,1,...\label{eqn16}\end{eqnarray}
 can be generated from the n-th order approximation by implementing
the following algorithm:- Use $g_{ab}^{(n)}$ to define $\Delta_{ab}^{(n)}$
and $\Box^{(n)}$ from (\ref{eqn9}) and (\ref{eqn10}) respectively,
and then use (\ref{eqn12}) and (\ref{eqn11}) with (\ref{eqn15})
and (\ref{eqn16}) to define the recursive relations \begin{eqnarray}
\Box^{(n)}{\mathcal{M}}^{(n+1)} & = & \Box^{(n)}\left[{\mathcal{M}}^{(n)}+E^{(n+1)}\right]=4\label{eqn17}\\
g_{ab}^{(n+1)} & = & g_{ab}^{(n)}+\epsilon_{ab}^{(n+1)}=\Delta_{ab}^{(n)}{\mathcal{M}}^{(n+1)};\,\, n=0,1,...\label{eqn18}\end{eqnarray}
 Since ${\mathcal{M}}^{(0)}$ represents the Minkowski vacuum and
$g_{ab}^{(0)}=\gamma_{ab}$ by (\ref{eqn13}) and (\ref{eqn14}),
then (\ref{eqn17}) together with (\ref{eqn18}) can be used to generate
a sequence of successive approximations to the exact solution.

In particular, since it is easily shown that $\Box^{(0)}{\mathcal{M}}^{(0)}=4$,
then it follows from (\ref{eqn17}) that $\Box^{(0)}E^{(1)}=0$. Consequently,
for a spherically symmetric disturbance, we find the retarded solution\[
E^{(1)}=\frac{H(R-ct)}{R}\]
 so that, as a first approximation, we have\begin{equation}
{\mathcal{M}}^{(1)}={\mathcal{M}}^{(0)}+\frac{H(R-ct)}{R}.\label{eqn18a}\end{equation}

\section{The Absence of Dipole Effects.}

\label{appC} Putting $n=1$ in (\ref{eqn17}) leads to the equation
for the second order approximation, and this is given by \[
\Box^{(1)}{\mathcal{M}}^{(2)}\equiv\Box^{(1)}\left({\mathcal{M}}^{(1)}+E^{(2)}\right)=4\]
 so that \begin{equation}
\Box^{(1)}E^{(2)}=4-\Box^{(1)}{\mathcal{M}}^{(1)}.\label{eqn20}\end{equation}
 \par{}

Our analysis proceeds in the following way:- by showing that the source
term in this equation has the form of an octopole multiplied by a
regular function, we are able to show that $E^{(2)}$ cannot be of
lower order than a quadropole. \par{}

The analysis requires that we consider the explicit form of the operator
$\Box^{(1)}$ which is the first order approximation of the general
wave operator \begin{equation}
\Box\equiv g^{ij}\left(\frac{\partial^{2}.}{\partial x^{i}\partial x^{j}}-\Gamma_{ij}^{k}\frac{\partial.}{\partial x^{k}}\right)\label{eqn21}\end{equation}
 where \[
\Gamma_{ij}^{k}=\frac{1}{2}g^{kr}\left(\frac{\partial g_{jr}}{\partial x^{i}}+\frac{\partial g_{ri}}{\partial x^{j}}-\frac{\partial g_{ij}}{\partial x^{r}}\right).\]
 This approximation is obtained using \begin{equation}
g_{ab}^{(1)}=\gamma_{ab}+\epsilon_{ab}^{(1)}\label{eqn22}\end{equation}
 which, for small disturbances, implies \[
g_{(1)}^{ab}\approx\gamma_{ab}-\epsilon_{ab}^{(1)},\]
 where the disturbance is given by \begin{equation}
\epsilon_{ab}^{(1)}=\frac{\partial^{2}E^{(1)}}{\partial x^{a}\partial x^{b}}=\frac{\partial^{2}}{\partial x^{a}\partial x^{b}}\left(\frac{H(R-ct)}{R}\right),\label{eqn23}\end{equation}
 and satisfies \begin{equation}
\gamma_{ij}\epsilon_{ij}^{(1)}\equiv\gamma_{ij}\frac{\partial^{2}E^{(1)}}{\partial x^{i}\partial x^{j}}=0.\label{eqn24}\end{equation}

Substitution of (\ref{eqn22}) into the definition (\ref{eqn21})
of $\Box$ gives \begin{equation}
\Box^{(1)}\approx\left(\gamma_{ij}-\epsilon_{ij}^{(1)}\right)\left(\frac{\partial^{2}.}{\partial x^{i}x^{j}}-\Gamma_{ij}^{k}\frac{\partial.}{\partial x^{k}}\right)\label{eqn25}\end{equation}
 where \begin{equation}
\Gamma_{ij}^{k}=\frac{1}{2}\left(\gamma_{kr}-\epsilon_{kr}^{(1)}\right)\left(\frac{\partial\epsilon_{jr}^{(1)}}{\partial x^{i}}+\frac{\partial\epsilon_{ri}^{(1)}}{\partial x^{j}}-\frac{\partial\epsilon_{ij}^{(1)}}{\partial x^{r}}\right).\label{eqn26}\end{equation}

Consequently, \begin{equation}
\Box^{(1)}=\Box^{(0)}-\gamma_{ij}\Gamma_{ij}^{k}\frac{\partial.}{\partial x^{k}}-\epsilon_{ij}^{(1)}\frac{\partial^{2}.}{\partial x^{i}\partial x^{j}}+\epsilon_{ij}^{(1)}\Gamma_{ij}^{k}\frac{\partial.}{\partial x^{k}},\label{eqn27}\end{equation}
 where $\Box^{(0)}$ is the flat space-time wave operator. The second
term on the right of this expression is identically zero; to see this,
we first note that, by (\ref{eqn26}), it contains the factor \begin{equation}
\gamma_{ij}\left(\frac{\partial\epsilon_{jr}^{(1)}}{\partial x^{i}}+\frac{\partial\epsilon_{ri}^{(1)}}{\partial x^{j}}-\frac{\partial\epsilon_{ij}^{(1)}}{\partial x^{r}}\right).\label{eqn28}\end{equation}
 \par{}

Using the definition (\ref{eqn23}) of $\epsilon_{ab}^{(1)}$, we
find \[
\gamma_{ij}\frac{\partial\epsilon_{jr}^{(1)}}{\partial x^{i}}=\gamma_{ij}\frac{\partial}{\partial x^{i}}\left(\frac{\partial^{2}E^{(1)}}{\partial x^{j}\partial x^{r}}\right)=\frac{\partial}{\partial x^{r}}\left(\gamma_{ij}\frac{\partial^{2}E^{(1)}}{\partial x^{i}\partial x^{j}}\right).\]
 Using (\ref{eqn24}) we immediately obtain \[
\gamma_{ij}\frac{\partial\epsilon_{jr}^{(1)}}{\partial x^{i}}=0.\]
 A similar analysis of the remaining terms in the expression (\ref{eqn28})
gives \[
\gamma_{ij}\frac{\partial\epsilon_{ri}^{(1)}}{\partial x^{j}}=\gamma_{ij}\frac{\partial\epsilon_{ij}^{(1)}}{\partial x^{r}}=0,\]
 so that, finally, \[
\gamma_{ij}\Gamma_{ij}^{k}=0,\]
 for the result. \par{}

We now consider the fourth term of (\ref{eqn27}):- by (\ref{eqn26})
this contains the factor \begin{equation}
\epsilon_{ij}^{(1)}\left(\frac{\partial\epsilon_{jr}^{(1)}}{\partial x^{i}}+\frac{\partial\epsilon_{ri}^{(1)}}{\partial x^{j}}-\frac{\partial\epsilon_{ij}^{(1)}}{\partial x^{r}}\right).\label{eqn29}\end{equation}
 By (\ref{eqn19}) the first term of (\ref{eqn29}) can be expressed
as \[
\epsilon_{ij}^{(1)}\frac{\partial\epsilon_{jr}^{(1)}}{\partial x^{i}}=\left(\frac{\ddot{H}}{R}S_{i}S_{j}+\frac{\dot{H}}{R^{2}}\left(RS_{ij}-R_{i}S_{j}-R_{j}S_{i}\right)-\frac{H}{R^{3}}\left(RR_{ij}-2R_{i}R_{j}\right)\right)\times\]
 \[
\frac{\partial}{\partial x^{i}}\left(\frac{\ddot{H}}{R}S_{j}S_{r}+\frac{\dot{H}}{R^{2}}\left(RS_{jr}-R_{j}S_{r}-R_{r}S_{j}\right)-\frac{H}{R^{3}}\left(RR_{jr}-2R_{j}R_{r}\right)\right)\]
 \[
=\left(\frac{\ddot{H}}{R}S_{i}S_{j}+\frac{\dot{H}}{R^{2}}\left(RS_{ij}-R_{i}S_{j}-R_{j}S_{i}\right)-\frac{H}{R^{3}}\left(RR_{ij}-2R_{i}R_{j}\right)\right)\times\]
 \[
\left(\frac{\dot{H}}{R}S_{i}S_{j}S_{r}-\frac{\ddot{H}}{R^{2}}\left(R_{i}S_{j}S_{r}+S_{i}R_{j}S_{r}+S_{i}S_{j}R_{r}\right)+\frac{\ddot{H}}{R}\left(S_{ij}S_{r}+S_{j}S_{ir}+S_{i}S_{jr}\right)\right.\]
 \[
-\frac{\dot{H}}{R^{2}}\left(S_{jr}S_{i}+S_{jr}R_{i}+S_{ij}S_{r}+S_{ij}R_{r}+S_{ir}S_{j}+S_{ir}R_{j}\right)\]
 \[
+\left.\frac{2\dot{H}}{R^{3}}\left(S_{i}R_{j}R_{r}+S_{j}R_{i}R_{r}+S_{r}R_{i}R_{j}\right)+\frac{\dot{H}}{R}S_{ijr}+O\left(\frac{1}{R^{4}}\right)\right).\]
 Use of the easily proven identities \begin{eqnarray}
S_{k}S_{k} & \equiv & 0,\nonumber \\
S_{k}R_{k} & = & 1,\nonumber \\
S_{k}S_{kj} & \equiv & 0\label{eqn30}\\
S_{ij}S_{ij} & = & \frac{2}{R^{2}}\nonumber \\
S_{abc} & = & -\frac{R_{a}}{R}S_{bc}-\frac{R_{b}}{R}S_{ca}-\frac{R_{c}}{R}S_{ab}\nonumber \end{eqnarray}
 in this expression leads to \[
\epsilon_{ij}^{(1)}\frac{\partial\epsilon_{jr}^{(1)}}{\partial x^{i}}\approx P_{4}\times S_{r},\]
 where $P_{4}$ denotes an octopole term. Similar analyses on the
remaining two terms of (\ref{eqn29}) then gives the result \[
\epsilon_{ij}^{(1)}\Gamma_{ij}^{k}\approx P_{4}\times S_{k}\]
 \par{}

Thus, we can finally write \[
\Box^{(1)}\approx\Box^{(0)}-\epsilon_{ij}^{(1)}\frac{\partial^{2}.}{\partial x^{i}\partial x^{j}}+P_{4}\times S_{k}\frac{\partial.}{\partial x^{k}},\]
 so that (\ref{eqn20}) can be written as \[
\Box^{(0)}E^{(2)}-\epsilon_{ij}^{(1)}\frac{\partial^{2}E^{(2)}}{\partial x^{i}\partial x^{j}}+P_{4}\times S_{k}\frac{\partial E^{(2)}}{\partial x^{k}}=\]
\begin{equation}
4-\Box^{(0)}{\mathcal{M}}^{(1)}+\epsilon_{ij}^{(1)}\frac{\partial^{2}{\mathcal{M}}^{(1)}}{\partial x^{i}\partial x^{j}}-P_{4}\times S_{k}\frac{\partial{\mathcal{M}}^{(1)}}{\partial x^{k}}.\label{eqn31}\end{equation}
 Putting n=0 in (\ref{eqn17}) shows that the first two terms on the
right side of this expression cancel, so let us now consider the third
term:- From the first order solution we have \[
\frac{\partial^{2}{\mathcal{M}}^{(1)}}{\partial x^{a}\partial x^{b}}\equiv g_{ab}^{(1)}=\gamma_{ab}+\epsilon_{ab}^{(1)}\]
 so that, for the third term, we get \[
\epsilon_{ij}^{(1)}\frac{\partial^{2}{\mathcal{M}}^{(1)}}{\partial x^{i}\partial x^{j}}=\epsilon_{ij}^{(1)}\gamma_{ij}+\epsilon_{ij}^{(1)}\epsilon_{ij}^{(1)}.\]
 Since $\gamma_{ij}\epsilon_{ij}^{(1)}=0$ by (\ref{eqn24}), then
we need only consider the quadratic term. Using (\ref{eqn19}) and
the relations (\ref{eqn30}), we can easily show that this leads to
the result \[
\epsilon_{ij}^{(1)}\epsilon_{ij}^{(1)}\approx P_{4}.\]
 Finally, we consider the fourth term on the right of (\ref{eqn31}):-
since the first order solution is given by \[
{\mathcal{M}}^{(1)}={\mathcal{M}}^{(0)}+\frac{H(R-ct)}{R},\]
 where $U^{(0)}$ is a regular scalar function, then this final term
of (\ref{eqn31}) becomes \[
P_{4}\times S_{k}\frac{\partial{\mathcal{M}}^{(1)}}{\partial x^{k}}=P_{4}\left(\frac{\partial{\mathcal{M}}^{(0)}}{\partial x^{k}}S_{k}+\frac{\dot{H}}{R}S_{k}S_{k}-\frac{H}{R^{2}}S_{k}R_{k}\right).\]
 Since $S_{k}$ is a vector with regular components (see (\ref{eqn19})),
and $U^{(0)}$ is a regular scalar which is everywhere twice differentiable,
then the first term in the brackets on the right of this expression
is also a regular scalar function; use of the identities $S_{k}S_{k}=0$
and $S_{k}R_{k}=1$ then gives \[
P_{4}\times S_{k}\frac{\partial{\mathcal{M}}^{(1)}}{\partial x^{k}}=P_{4}\times\left(G(\mathbf{x})+P_{2})\right)\approx P_{4}\times G(\mathbf{x}),\]
 for some regular function $G$, and where $P_{2}$ denotes a dipole
term. Collecting these results gives, from (\ref{eqn31}), \begin{equation}
\Box^{(0)}E^{(2)}-\epsilon_{ij}^{(1)}\frac{\partial^{2}E^{(2)}}{\partial x^{i}\partial x^{j}}+P_{4}\times S_{k}\frac{\partial E^{(2)}}{\partial x^{k}}\approx P_{4}\times{\mathcal{M}}(\mathbf{x}).\label{eqn32}\end{equation}
 Remembering that, according to (\ref{eqn19}), $\epsilon_{ab}^{(1)}$
behaves as a monopole it is now simple to show that $E^{(2)}$ must
have the general form \[
E^{(2)}=\frac{G(R-ct)}{R^{N}}\times(Regular\, function);~~N>2.\]
 Consequently, the second order approximation to the stellar model
can be expressed as \[
{\mathcal{M}}^{(2)}(\mathbf{x})\approx{\mathcal{M}}^{(0)}(\mathbf{x})+\frac{H(R-ct)}{R}+\frac{G(R-ct)}{R^{N}}\times(Regular\, function)+...,\]
 where $N>2$, so that, according to the vacuum gravity formalism,
dipole effects are absent in an arbitrarily defined spherically symmetric
disturbance of the Minkowski vacuum.

\end{document}